# Exploiting Web Service Semantics: Taxonomies vs. Ontologies

Asuman Dogac, Gokce Laleci, Yildiray Kabak, Ibrahim Cingil

Software Research and Development Center
Middle East Technical University (METU)
06531 Ankara Turkiye
email: asuman@srdc.metu.edu.tr

**Abstract**

*Comprehensive semantic descriptions of Web services are essential to exploit them in their full potential, that is, discovering them dynamically, and enabling automated service negotiation, composition and monitoring. The semantic mechanisms currently available in service registries which are based on taxonomies fail to provide the means to achieve this. Although the terms "taxonomy" and "ontology" are sometimes used interchangably there is a critical difference. A taxonomy indicates only class/subclass relationship whereas an ontology describes a domain completely. The essential mechanisms that ontology languages provide include their formal specification (which allows them to be queried) and their ability to define properties of classes. Through properties very accurate descriptions of services can be defined and services can be related to other services or resources.*

*In this paper, we discuss the advantages of describing service semantics through ontology languages and describe how to relate the semantics defined with the services advertised in service registries like UDDI and ebXML.*

## 1  Introduction

When looking towards the future of web-services, it is predicted that the breakthrough will come when the software agents start using web-services rather than the users who need to browse to discover the services. Currently well accepted standards like Web Services Description Language [WSDL] and Simple Object Access Protocol [SOAP] make it possible only to "dynamically access" to Web services in an application. That is, when the service to be used is known, its WSDL description can be accessed by a program which uses the information in the WSDL description like the interface, binding and operations to dynamically access the service. However to dynamically *discover* services, say through software agents require detailed semantic information about the services to be available.

Currently, a number of taxonomies are being used to discover services in service registries like [UDDI] or [ebXML]. The most widely used taxonomies are North American Industrial Classification Scheme [NAICS] for associating services with "industry" semantics; Universal Standard Products and Services Classification [UNSPSC] for classifying product/services and [ISO 3166] for locale.